\documentstyle[12pt]{article}
\def\sign{{\rm sign}\;}
\def\const{{\rm const}}
\def\diag{{\rm diag}\;}
\def\e{{\rm e}}
\def\to{\rightarrow}
\def\sumi{\sum_{i=1}^{n}}
\begin{document}

\begin{center}
               RUSSIAN GRAVITATIONAL ASSOCIATION\\
               CENTER FOR SURFACE AND VACUUM RESEARCH\\
               DEPARTMENT OF FUNDAMENTAL INTERACTIONS AND METROLOGY
\end{center}
\vskip 4ex
\begin{flushright}                           RGA-CSVR-003/94\\
                                             gr-qc/9403063
\end{flushright}
\vskip 45mm

\begin{center}
     {\bf THE BIRKHOFF THEOREM IN MULTIDIMENSIONAL GRAVITY \\
\vskip2.5ex
     K.A.Bronnikov and V.N.Melnikov}\\
\vskip 5mm
     {\em Centre for Surface and Vacuum Research,\\
     8 Kravchenko str., Moscow, 117331, Russia}\\
     e-mail: mel@cvsi.uucp.free.msk.su\\
\vskip 60mm

             Moscow 1994
\end{center}
\pagebreak

\setcounter{page}{1}

\begin{center}
     {\Large\bf THE BIRKHOFF THEOREM \\ IN MULTIDIMENSIONAL GRAVITY \\
\vskip2.5ex
     K.A.Bronnikov and V.N.Melnikov}\\
\vskip2.5ex
     {\em Center for Surface and Vacuum Research,\\
     8 Kravchenko str., Moscow, 117331, Russia}\\
     e-mail: mel@cvsi.uucp.free.msk.su
\end{center}
\vskip2ex

{\bf ABSTRACT}

\bigskip
\noindent
     The validity conditions for the extended Birkhoff theorem in
     multidimensional gravity with $n$ internal spaces are formulated,
     with no restriction on space-time dimensionality and signature.
     Examples of matter sources and geometries for which the theorem is
     valid are given. Further generalization of the theorem is discussed.
\vskip3ex

\section{INTRODUCTION}

     The original Birkhoff theorem [1] states that in general relativity
     (GR) the spherically symmetric vacuum field is static and is thus
     reduced to the Schwarzschild solution. From a wider viewpoint, the
     theorem indicates a case when the field equations induce, under certain
     circumstances, an additional field system symmetry that was not
     postulated at the outset.  The theorem is closely related to the
     quadrupole nature of the gravitational field in GR, more precisely, to
     the absence of monopole gravitational waves. Thus theorems of this sort
     are able not only to simplify the treatment of certain physically
     relevant situations but also to provide their better understanding.

     After Birkhoff the theorem was extended to spherical systems with a
     nonzero cosmological constant $\Lambda$, the Maxwell or Born-Infeld
     electromagnetic fields (\cite{`2,`3} and others), scalar fields and
     $\Lambda\ne 0$ in GR \cite{`4} and some scalar-tensor theories of
     gravity \cite{`5}.  In Ref. \cite{`6} the theorem was extended to
     planarly and pseudospherically symmetric Einstein-Maxwell fields.

     Another approach was suggested in Refs. \cite{`7,`8}: the study was
     aimed at finding out general conditions under which the staticity
     theorem could be proved. This allowed all the previously found cases of
     GR and scalar-tensor theories when the extended Birkhoff theorem is
     valid, to be included, along with many new ones. The theorem was
     generalized in two respects: to include more types of space-time
     symmetry (e.g., planar, cylindrical and pseudoplanar) and more kinds of
     matter (scalar fields, gauge fields, perfect fluid, etc.).

     Here we would like to extend the approach of \cite{`7,`8} to
     multidimensional gravity. One may recall that most modern unification
     theories incorporate more than four dimensions (e.g., that of
     superstrings \cite{`9}); on the other hand, some studies are undertaken
     in (2+1) and even (1+1) dimensions where certain hard problems simplify
     and admit a deeper insight. The low energy
     limit of many theories, actually embracing an enormous range of energy
     scales, is reduced to the multidimensional Einstein equations
\begin{equation}                                                    
     G_A^B\equiv R_A^B - \delta_A^B R_C^C/2= -T_A^B,
\end{equation}
     where $R_A^B$ is the $D$-dimensional Ricci
     tensor and $T_A^B$ is the matter energy-momentum tensor (EMT). We will
     assume the validity of (1) for some dimension $D$ and some kind of
     matter and find certain general conditions under which these field
     equations make the system symmetry increase. The consideration
     essentially follows the lines of \cite{`7,`8}.  In Section 2 the
     extended Birkhoff theorem is proved for multidimensional GR. In Section
     3 its different special cases are discussed and
     Section 4 contains some remarks, in particular, on situations
     excluded by the requirements of the theorem; its further
     extension to multidimensional scalar-tensor theories is presented.

\section{THEOREM}

     Consider a $D$-dimensional Riemannian or pseudo-Riemannian space with
     the structure
\begin{equation}                                                      
     V^D = M^2\times V_1\times V_2 \times \ldots \times V_n,\ \
     {\rm dim}\,V_i=N_i,\ \ n=1,2,\ldots
\end{equation}
     where $M^2$ is an arbitrary two-dimensional subspace parametrized by the
     coordinates $u$ and $v$ and $V_i$ are subspaces of arbitrary dimension
     ($N_i$) and signature whose metric depends on $u$ and $v$ only via
     conformal (scale) factors. Thus with no further loss of generality the
     $D$-dimensional metric may be written in the form
\begin{equation}                                                        
     ds_D^2 = \eta_u \e^{2\alpha}du^2 +
     \eta_v \e^{2\gamma}dv^2 + \sumi\e^{2\beta_i}ds_i^2
\end{equation}
     where $\eta_u=\pm 1,\ \eta_v= \pm 1$; $\alpha,\ \beta_i$ and $\gamma$
     are functions of $u$ and $v$ and $ds_i^2$ are the $u$- and
     $v$-independent metrics of the subspaces. It is meant that $M^2$ along
     with one (two-dimensional) or two (one-dimensional) subspaces $V_i$ form
     the conventional physical space-time while the rest $V_i$ correspond to
     extra (internal) dimensions. For greater generality we would not like to
     fix the signs $\eta_u$ and $\eta_v$.

     Before formulating the theorem let us introduce the quantity
\begin{equation}
     \rho(u,v) \equiv \sumi N_i\beta_i(u,v).              \label{rho}
\end{equation}
     and present the nonzero Ricci tensor components for the metric (3):
\begin{eqnarray}
     R_u^u &=& \Box_u \gamma+ \Box_v \alpha +
          \eta_u\e^{-2\alpha}
          \bigl(\rho''-\alpha'\rho'+\sumi \beta_i^{'2}\bigr) +
          \eta_v\e^{-2\gamma}\dot\alpha \dot\rho;           \label{Ruu}\\
     R_v^v &=& \Box_u \gamma+ \Box_v \alpha +
          \eta_u\e^{-2\alpha}\gamma'\rho' +
          \eta_v\e^{-2\gamma}
          \bigl(\ddot\rho-\dot\gamma\dot\rho +\sumi \dot\beta_i^2\bigr);
                                                            \label{Rvv}\\
     R^{m_i}_{n_i} &=& \e^{-2\beta_i}\overline{R}^{m_i}_{n_i} +
          \delta^{m_i}_{n_i}\bigl[(\Box_u + \Box_v)\beta_i +
          \eta_u\e^{-2\alpha}\beta'_i\rho' +
          \eta_v\e^{-2\gamma}\dot\beta_i\dot\rho\bigr];
                                                         \label{Rmn}\\[1.2ex]
     R_{uv} &=& \dot\rho' - \gamma'\dot\rho -\dot\alpha \rho'+
          \sumi \dot\beta_i \beta'_i                        \label{Ruv}
\end{eqnarray}
     where primes and dots stand for partial derivatives $\partial_u$ and
     $\partial_v$, respectively, and
\begin{equation}
 \Box_u=\e^{-\alpha-\gamma}\partial_u(\e^{\gamma-\alpha}\partial_u),\ \
     \Box_v=\e^{-\alpha-\gamma}\partial_v(\e^{\alpha-\gamma}\partial_v),
\end{equation}
     The indices $m_i$ and $n_i$ belong to the subspace $V_i$; the Ricci
     tensors $R^{m_i}_{n_i}$ correspond to the metrics $ds_i^2$ and do not
     depend on $u$ and $v$.
\medbreak
\noindent {\bf Theorem 1.} \ \
     {\sl Let there be a Riemannian space $V^D$ {\rm (2)} with metric {\rm
     (3)} obeying the Einstein equations {\rm (1)}. If}
\vskip-1.0ex
\begin{description}
\item{{\bf (A)}} {\sl there is a domain $\Delta$ in $M^2$ where}
\begin{equation}
                \sign(\rho^{,A}\rho_{,A})=\eta_u;         \label{signrho}
\end{equation}
\vskip-1.0ex
\item{{\bf (B)}} {\sl each $\beta_i(u,v)$ in $\Delta$ is functionally
     related to $\rho$} {\rm (certain relations $F_i(\rho, \beta_i)=0$ are
     valid);}
\item{{\bf (C)}} {\sl in an orthogonal coordinate frame where}
     $\rho=\rho(u)$ {\rm (its existence is guaranteed by (10))}
     {\sl the EMT component $T_{uv}\equiv 0$ and there is a combination}
\begin{equation}
     T^v_v +{\rm const}\cdot T^u_u                          \label{Tcomb}
\end{equation}
\vskip-1.0ex
     {\sl independent of $v$ and $\gamma$,}
\end{description}
     {\sl then there is an orthogonal coordinate frame $(u,v)$ in $\Delta$
     such that the metric {\rm (3)} is $v$-independent.}
\medbreak

     {\it Proof.}\ Let us choose an orthogonal coordinate frame where
     $\rho=\rho(u)$, which is possible by Condition A. Then by Condition B
     all $\dot\beta_i=0$. By Condition C the mixed EMT
     component $T_{uv}=0$ (in the conventional case $\eta_u=-\eta_v$ that
     means that there is no energy flow in the frame of reference where
     $\rho=\rho(u)$), and the corresponding component of Eqs.(1) yields
     $\dot\alpha =0$. Now only $\gamma$ may depend on $v$. To make the
     last step and to obtain $\gamma=\gamma(u)$ it is sufficient to find a
     combination of the Einstein equations having the form $\gamma'=f(u)$,
     whence
\begin{equation}
     \gamma=\gamma_1(u) + \gamma_2(v)                       \label{Gamma}
\end{equation}
     and $\gamma_2$ may be brought to zero
     by a coordinate transformation $v=v(\tilde v)$. Observing
     (5-7), one can see that any combination of the form
     $G_u^u + {\rm const}\cdot G_v^v$ of the left-hand sides of (1)
     does contain $\gamma$ but only in the term $\e^{-2\alpha}\rho'\gamma'$.
     As $\rho\ne\const$, our problem is solved when the corresponding
     combination of $T_A^B$ does not depend on $\gamma$ and $v$, exactly what
     is required in Condition C. This completes the proof.
\medbreak

     The theorem generalizes the results of \cite{`7,`8} to arbitrary
     space-time dimension and signature, including multidimensional
     Kaluza-Klein-type models with a chain of internal spaces each with a
     scale factor of its own, such as considered in, e.g., [11, 12].

\section{SPECIAL CASES}

     In the following examples, unless otherwise indicated, we will adhere to
     the conventional interpretation of the Birkhoff theorem, i.e., assume
     that $v$ is time ($\eta_v=1$ and $u$ is a space variable
     ($\eta_u=-1$). Everything may be easily reformulated for
     coinciding $\eta_u$ and $\eta_v$. No assumptions on the signatures of
     $V_i$ are made since they do not affect the conclusions.

     In general, the following matter sources satisfy the requirements C of
     Theorem 1 with no further restrictions on the structure of $V^D$:
\begin{description}
\item[(a)]
     Linear or nonlinear, minimally coupled scalar fields with the
     Lagrangian $L=\varphi^{,A}\varphi_{,A}-V(\varphi)$ where $V(\varphi)$ is
     an arbitrary function, under the restriction $\varphi=\varphi(u)$:
\begin{equation}
     2T^A_B = \delta^A_B V(\varphi) + \eta_u\e^{-2\alpha}{\varphi'}^2
                \diag (1,\ -1,\ \ldots,\ -1);
                                                                \label{eqS}
\end{equation}
     here and henceforth positions in ``$\diag$'' are ordered by the scheme
     $(u,\ v,\ \ldots)$.
\item[(b)]
     A massless, minimally coupled scalar field
     ($L= \varphi^{,A}\varphi_{,A})$ under the restriction
     $\varphi=\varphi(v)$: the EMT does not contain $\varphi$ but only
     $\dot\varphi = \const$ (the so-called cosmological scalar field):
\begin{equation}
     2T^A_B = \eta_v\e^{-2\gamma}\to\dot\varphi^2
                \diag (-1,\ 1,\ -1,\ \ldots,\ -1).
                                                                \label{eqST}
\end{equation}
\item[(c)]
     Abelian gauge fields ($L=-F^{AB}F_{AB},\
     F_{AB}= \partial_A U_B - \partial_B U_A$) under the restriction that the
     vector potential $U_A$ has a single nonzero component $U_K (u)$, with
     a fixed coordinate $K\ne v$, so that among $F_{AB}$ only
     $F_{uK}=-F_{Ku}\ne 0$:
\begin{equation}
       T^u_u= T^K_K =-F^{uK}F_{uK};\ \ \
      {\rm other}\ \  T^A_B= \delta^A_B F^{uK}F_{uK}.
\label{eqE}
\end{equation}
\item[(d)]
     Nonlinear vector fields with Lagrangians of the form $\Phi(I),\
     I = F^{AB}F_{AB}$,
     where $\Phi$ is an arbitrary function, under the same restriction as
     that in item (c) but with $K=v$ (an example is the Born-Infeld nonlinear
     electromagnetic field):
\begin{equation}
     T^u_u = T^v_v = 2(d\Phi/dI)F^{uv}F_{uv}-\Phi/2;\ \ \
      {\rm other}\ \  T^A_B= -\delta^A_B \Phi/2.
\label{eqEN}
\end{equation}
\item[(e)] Some kinds of interacting fields: for instance, the system of an
     Abelian gauge field and a scalar dilaton field
     ($L=\varphi^{,A}\varphi_{,A} - \e^{2\lambda\varphi}F^{AB}F_{AB},\
     \lambda=\const$) under the constraints of items (a) and (c): the EMT
     structure combines those of (13) and (15). As (\ref{eqS}) is
     $v$- and $\gamma$-independent, evidently the second condition C of
     Theorem 1 is satisfied by one of the two combinations $T^u_u \pm T^v_v$.
     This is just the interaction relevant for multidimensional dilatonic
     black holes [11-13, 16].

     The same is true if the expression $\e^{2\lambda\varphi}$ in the
     Lagrangian is replaced by any function of $\varphi$.
\item[(f)] The cosmological term $\Lambda\delta^A_B$ may be added to the
     left-hand side of (1) with no consequences.
\end{description}

     One can easily find other forms of matter, as well as combinations of
     the above forms of matter and other ones, for which Theorem 1 holds.

     As for the diversity of space-time structures to which the theorem
     applies, it is also very wide. In the 4-dimensional case it includes the
     symmetries mentioned in \cite{`7,`8}, namely: spherical, planar,
     pseudospherical, pseudoplanar, cylindrical, toroidal ($V^D= M^2\times
     S^2,\ \ M^2\times R^2,\ \ M^2\times L^2,\ \ M^2\times R^1\times R^1, \
     \ M^2\times R^1\times S^1,\ \  M^2\times S^1\times S^1$, respectively,
     where $L^2$ is the Lobachevsky plane). It applies to both conventional
     (Lorentzian) GR and its ``Euclidean'' counterpart, as well as to
     Kaluza-Klein type models with a chain of internal spaces with
     $u$-dependent scale factors. Moreover, multidimensional extensions may
     incorporate generalized spherical and other symmetries in the spirit
     of Tangherlini \cite{Tan}, i.e., $S^m,\ R^m$ or $L^m$ with an
     arbitrary $m>2$ instead of $S^2,\ R^2,\ L^2$.

     For space-times with horizons, such as the black-hole ones, the theorem
     states the metric independence on different coordinates in different
     domains of $M^2$: thus, in the conventional Schwarzschild case it fixes
     the  $t$-independence (staticity) in the $R$ domain and  $r$-independence
     (homogeneity) in the $T$ domain. The same applies to multidimensional
     black holes considered in many papers (e.g., [11-16]).

     Another point of interest is the existence of Abelian gauge fields
     of various directions which satisfy the theorem, see the above item (c).
     One may recall such evident examples as Coulomb-like fields for
     spherical and other similar symmetries, radial, longitudinal and
     azimuthal electric fields for conventional cylindrically symemtric
     space-times (and their magnetic counterparts); however, there are
     configurations with $u$-dependent vector potential components directed
     in extra dimensions, deserving a separate treatment.

\section{COMMENTS}

\noindent{\bf 4.1.}
     Condition A of Theorem 1 may be weakened if $M^2$ is a proper Riemannian
     space ($\eta_u=\eta_v$): instead of (\ref{signrho}), it is sufficient
     to assume just $\rho\ne\const$. Indeed, in this case the orthogonal
     coordinates $u$ and $v$ may be always chosen so that $\rho=\rho(u)$,
     for instance, one may put just $u=\rho$.

     If $M^2$ is pseudo-Riemannian  ($\eta_u=-\eta_v$), then the gradient of
     $\rho(u,v)$ may be either $u$-like, or $v$-like, or null
     ($\sign(\rho^{,A}\rho_{,A})= \eta_u,\ \eta_v,\ 0 $, respectively).
     To extend the theorem to the case when it is $v$-like one may just
     change the notations of the coordinates, $u \leftrightarrow v$,
     irrespective of which of them is spacelike. So in both cases one
     can achieve $\rho=\rho(u)$.
\medbreak
\noindent{\bf 4.2.}
     Cancellation of Condition B of Theorem 1 (possible if $n>1$) leads
     to the existence of at least two functionally independent unknowns.
     The situation is most obviously exemplified by the Einstein-Rosen
     vacuum cylindrical gravitational waves \cite{ER} ($D=4,\ N_1=N_2=1,\
     v=t$, i.e., time).

     More generally, extra-dimension scale factors behave like minimally
     coupled scalar fields in 4 dimensions, so possible monopole waves may
     be eliminated just at the expense of the additional assumption B.
     When the latter is removed, these waves may manifest the instability
     of static configurations (as is the case for many non-black-hole
     spherically symmetric multidimensional space-times [11, 13].
 \medbreak
 \noindent{\bf 4.3.}
     The possibility $\rho=\const$, excluded in the theorem, looks somewhat
     exotic in the spherically symmetric case but is quite natural for, say,
     planar symmetry. Let us show that it leads to wave solutions to Eqs. (1)
     taking as an example a 4-dimensional vacuum space-time with a
     cosmological constant ($T^A_B = \delta^A_B \Lambda$), possessing
     spherical or planar symmetry ($D=4,\ n=1,\ N_1=2;\ \overline{R}^2_2=
     \overline{R}^3_3= \epsilon = +1,\ 0$, respectively). In this case
     $\rho=2\beta_1$ and Eqs. (1) give:
\begin{equation}
     \epsilon\e^{-\rho}=\Lambda;\ \ \
          \e^{-2\alpha}(\eta_u\alpha''+ \eta_v\ddot\alpha) = \Lambda
                                                       \label{eq-const}
\end{equation}
     where the coordinates are chosen so that the metric of $M^2$ is
     conformally flat ($\alpha=\gamma$). There are the following variants:
\begin{itemize}
\item $\epsilon=1,\ \eta_u=-\eta_v;\ \Lambda>0$. The space-time is formed
     by a congruence of spheres of equal radii and the only nontrivial metric
     coefficient $\alpha$ obeys a nonlinear wave equation.
\item $\epsilon=1,\ \eta_u=\eta_v;\ \Lambda>0$. The same but the equation
     is nonlinear, elliptic type.
\item $\epsilon=0,\ \eta_u=-\eta_v;\ \Lambda=0$. Linear waves in a planarly
     symmetric space-time: $\alpha=\gamma=f_1(u+v)+f_2(u-v)$.
\item $\epsilon=0,\ \eta_u=\eta_v;\ \Lambda=0$. The only nontrivial metric
     coefficient $\alpha=\gamma$ is a harmonic function of $u$ and $v$.
\end{itemize}
\medbreak
\noindent{\bf 4.4.}
     Another possibility rejected in Theorem 1 is that $\rho(u,v)$
     has a null gradient in a pseudo-Riemannian $M^2$. The condition
     $\rho^{,A}\rho_{,A}=0$ in the coordinates such that $\alpha=\gamma$ leads
     to $\dot\rho = \pm \rho'$. Let us choose the plus sign (re-defining
     $u\to -u$ if required), so that $\rho=\rho(\xi),\ \xi=u+v$.

     Consider for instance vacuum planarly symmetric space-times of any
     dimension $D$, so that
\begin{equation}
     V^D= M^2\times R^{D-2};\ \ \ \rho=(D-2)\beta;\ \ \  T^A_B =0.
                                                                 \label{eq*}
\end{equation}
     Substituting $\rho=\rho(\xi)$ and $\alpha=\gamma$ for (\ref{eq*}) to
     Eqs.(1), we obtain:
\begin{equation}
     \alpha= \alpha_1(\xi)+\alpha_2(\eta);\ \ \
          2\alpha_1(\xi) = \ln|\beta'| + \beta,\ \ \beta=\beta(\xi)
                                                             \label{eq**}
\end{equation}
     where $\eta= u-v$; $\beta(\xi)$ and $\alpha_2(\eta)$ are arbitrary
     functions. This is a planarly symmetric vacuum wave solution to Eqs.
     (1) for any dimension $D$.

     The examples of items 4.3 and 4.4 show that in Theorem 1, establishing
     the sufficient conditions for staticity, no condition may
     be omitted or essentially weakened.
\medbreak
\noindent{\bf 4.5.}
     A case of interest is the one when some $\beta_i$ are only $u$-dependent
     while others linearly depend on $v$, so that $\dot\beta_i=$ const. If
     still $\dot\rho=0$ and the spaces $V_i$ corresponding to $\dot\beta_i
     \ne 0$ are Ricci-flat, then the proof of Theorem 1 may be properly
     modified to conclude that all the remaining metric coefficients are
     $v$-independent.

     Imagine, e.g., a 4-dimensional, static, spherically symmetric space-time
     ($u=r$, radial coordinate; $v=t$, time; $V_1=S^2$, a 2-dimensional
     sphere) accompanied by internal Ricci-flat spaces of which some are
     static (in general, $r$-dependent), others exponentially expanding
     ($\dot\beta_i = \const > 0$) and still others exponentially contracting
     ($\dot\beta_i =\const <0$). The specific forms of all functions are
     to be found from Eqs. (1) with a relevant choice of matter. However,
     as $\dot\rho=0$, this class of solutions cannot contain those in
     which all extra dimensions are contracting. (This certainly does not
     mean that such solutions cannot exist at all: they are just not
     covered by the present treatment.)
\medbreak
\noindent{\bf 4.6.}
     The consideration of Section 2 rests on
     the geometric structure of the space $V^D$, including certain symmetry
     requirements. However, the theorem contains no requirements to
     internal symmetries of the subspaces $V_i$ since no constraints on the
     dependence of $\overline{R}_{n_i}^{m_i}$ on the internal coordinates
     $y^{n_i}$ have appeared.  By Eqs. (1) this dependence is just
     the same as that in the EMT.

     However, in most applications, as seen from the examples of Section 3,
     $V_i$ are either Ricci-flat, or constant curvature spaces: as the EMT is
     independent of the internal coordinates, the same is true for $V_i$.
     Moreover, when all the diagonal components $T^{m_i}_{m_i}$ are equal
     to each other, Eqs.(1) force the components $\overline{R}^{m_i}_{m_i}$
     to be equal as well, while the off-diagonal components are zero. Thus
     it is the EMT symmetry that forces $V_i$ to be constant curvature
     spaces.

\medbreak
\noindent{\bf 4.7.}
     In \cite{`7,`8} the generalized
     Birkhoff theorem was extended to a broad class of scalar-tensor
     theories of gravity in 4 dimensions. The same can be done in the
     multidimensional case. To do that let us
     consider in $V^D$ with the metric $g_{MN}$ a scalar-tensor theory
     described by the Lagrangian
\begin{equation}
     \sqrt{g}L=\sqrt{g}\big[A(\phi)R + B(\phi)\phi^{,M}\phi_{,M}
          -2\Lambda(\phi) + L_m\big]
                                                            \label{STT}
\end{equation}
     where $g=\mid\det g_{MN}\mid; \ A>0,\ B$ and $\Lambda$ are any smooth
     functions of $\phi$ and the matter Lagrangian $L_m$ may depend on both
     $g_{MN}$ and $\phi$. The conformal mapping (suggested by Wagoner
     \cite{Wag} for $D=4$)
\begin{equation}
     g_{MN}= A^{-2/(D-2)}\overline{g}_{MN}                  \label{Conf}
\end{equation}
     brings (\ref{STT}) to the form (up to a divergence)
\begin{equation}
     \sqrt{g}L= \sqrt{\overline{g}}\bigg\{\overline{R}+
\frac{1}{A^2}\bigg[AB+\frac{D-1}{D-2}\left(\frac{dA}{d\phi}\right)^2\bigg]
     \overline{g}^{MN}\phi_{,M}\phi_{,N} +
     A^{-D/(D-2)}\big[-2\Lambda(\phi)+L_m\big]\bigg\}       \label{STT'}
\end{equation}
     where $\overline{g}=\mid\det\overline{g}_{MN}\mid$ and $\overline{R}$
     is the scalar curvature corresponding to $\overline{g}_{MN}$. Variation
     of (\ref{STT'}) with respect to $\overline{g}_{MN}$ yields the
     Einstein equations with an EMT containing the contribution of the
     (possibly nonlinear) scalar field $\phi$ and that of matter coupled
     to $\phi$. For our purpose it is essential that the latter
     contribution $\overline{T}_{MN}$ coincides with the original EMT
     $\bigl( T_{MN}= (\delta/\delta g^{MN})(\sqrt{g}L_m \bigr)$
     up to a $\phi$-dependent factor. Consequently, if $\phi=\phi(u)$
     and $T_{MN}$ satisfies Condition C of Theorem 1, so does
     $\overline{T}_{MN}$ and Theorem 1 is applicable to the metric
     $\overline{g}_{MN}$. However, now it is $\overline{\rho}=\rho+\ln A$
     that appears instead of $\rho$ in the formulation of the theorem.
     Therefore Theorem 1 cannot be directly applied to $g_{MN}$ and its
     formulation should be properly modified:

\medbreak
\noindent {\bf Theorem 2.} \ \
     {\sl Consider a field system with the Lagrangian {\rm (20)} in a
     Riemannian space $V^D$ {\rm (2)} with metric {\rm (3)}. Let there be a
     domain $\Delta$ in $M^2$ where}
\vskip-1.0ex
\begin{description}
\item[(i)]
     {\sl all $\beta_i$ and the field $\phi$ are functions of $u$;}
\vskip-1.7ex
\item[(ii)]
     $\overline{\rho}=\rho+\ln A \ne \const$ {\sl and}
\vskip-1.7ex
\item[(iii)]
     {\sl Conditions C of Theorem 1 are valid for the EMT derived from $L_m$.}
\end{description}
\vskip-1.0ex
     {\sl Then the coordinate $v$ in $\Delta$ may be chosen so that all
     $g_{MN}$ are $v$-independent.}

\medbreak
\noindent{\bf 4.8.}
     It would be of interest to try to extend the
     theorem to multidimensional models with nonzero off-diagonal metric
     components such as $g_{ui}$ with $i$ from extra dimensions, as is the
     case in the original Kaluza-Klein model. This goes beyond the
     scope of this paper, although probably such a
     generalization does exist since the new effective vector fields are
     unlikely to create monopole waves.

\vskip2ex
\centerline{\bf Acknowledgement}
\vskip2ex   \nopagebreak
     This work was supported in part by the Russian Ministry of Science.

\end{document}